\font\caps=cmcsc10 at 12pt
\font\eightrm=cmr8 at 8pt
\documentclass{oxarticle}
\usepackage{tikz}
\usepackage{amsthm}

\newtheorem{rant}{Rant}

\usetikzlibrary{backgrounds,fit,decorations.pathreplacing,calc}
\usepackage{amsmath, amssymb}
\newcommand{\cM}{{\cal M}}

\newcommand{\cG}{{\cal G}}

\newcommand{\cA}{{\cal A}}

\newcommand{\PB}{Master Equation}

\newcommand{\cL}{{\cal L}}
\newcommand{\bt}{\begin{tabular}{c}}
\newcommand{\et}{\end{tabular}}

\newcommand{\eb}{\ee\be } 
\newcommand{\ebp}{\rt.\ee\be\lt.} 
\newcommand{\bmat}{\lt ( \begin{array} }
\newcommand{\emat}{  \end{array} \rt )}

\newcommand{\oM}{{\ov M}}

\newcommand{\ovY}{{\ov \Psi}}

\newcommand{\om}{{\ov m}}

\newcommand{\oC}{{\ov C}}

\newcommand{\oF}{{\ov F}}
\newcommand{\A}{{\ov A}}

\renewcommand{\a}{\alpha}	
\renewcommand{\b}{\beta}
\newcommand{\g}{\gamma}
\renewcommand{\d}{\delta}
\newcommand{\e}{\epsilon}
\newcommand{\ve}{\varepsilon}

\newcommand{\lam}{\lambda}
\newcommand{\m}{\mu}
\newcommand{\n}{\nu}	
\newcommand{\x}{\xi}
	
\renewcommand{\r}{\rho}
\newcommand{\s}{\sigma}

\newcommand{\f}{\phi}
\renewcommand{\c}{\chi}

\newcommand{\G}{\Gamma}
\newcommand{\D}{\Delta}

\renewcommand{\P}{\Pi}

\newcommand{\Y}{\Psi} 

\newcommand{\W}{\Omega}
\newcommand{\la}{\label}
\newcommand{\ci}{\cite}

\newcommand{\ds}{\documentstyle}	
\newcommand{\fr}{\frac}

\newcommand{\pa}{\partial}
\newcommand{\ov}{\overline}
\newcommand{\br}{\begin{rant}}
\newcommand{\er}{\end{rant}}
\newcommand{\be}{\begin{equation}}
\newcommand{\ee}{\end{equation}}
\newcommand{\ba}{\begin{array}} 
\newcommand{\ea}{\end{array}}
\newcommand{\bea}{\begin{eqnarray}}
\newcommand{\eea}{\end{eqnarray}}
\newcommand{\ra}{\rightarrow}

\newcommand{\lra}{\longrightarrow}

\newcommand{\lt}{\left}
\newcommand{\rt}{\right}

\newcommand{\ben}{\begin{enumerate}}
\newcommand{\een}{\end{enumerate}}
\newcommand{\bitem}{\begin{itemize}}

\newcounter{orange} 
\newcounter{apple} 
\addtocounter{apple}{1}

\newcounter{grape} 
\addtocounter{grape}{1}
\newcommand{\numberhere}{JB277}
\newcommand{\articlenumber}{The1.5OrderisFlawed}

\proofmodefalse
\usepackage{color} 

\begin{document}
%\pagestyle{empty}

%\vspace*{1in}

\renewcommand{\thefootnote}{\fnsymbol{footnote}}
%\footnotetext[1]{~here we have a footnote.}\renewcommand{\thefootnote}\arabicfootnote}}

%\vfill
%
%

 \begin{center}
{\Huge { The 1.5 Order Formalism does not Generate\\ a Valid BRS Transformation for Supergravity}\\
[.5cm]} 
\vspace*{.1in}
{\caps John A. Dixon\footnote{cybersusy@gmail.com, \;john.dixon@ucalgary.ca}
\\ University of Calgary\\
Calgary, Alberta, Canada} \\[.5cm] 
\large
Abstract
\end{center}
\large
The 1.5 order formalism (sometimes called a `trick') is the cornerstone of modern supergravity. In this paper, the free massive Wess Zumino theory is used as a simple toy model to   look at the BRS symmetry of the first, second and  1.5 order formalisms.  This easily shows that the 1.5 order formalism is flawed for all theories.  The 1.5 algebra naively appears to close.  
However, when it is analyzed in detail, in a simple model,  where easy calculations are available, the 1.5 formalism always generates an invalid BRS operator, which is not even nilpotent. 
This clearly is also the case for supergravity.   It follows that  a revised and completed set of nilpotent first order supergravity transformations is needed to properly understand 3+1 dimensional supergravity.   Such a set  seems easy to write down,  by simply adding two more auxiliary fields so that the spin connection becomes part of a super--YM multiplet.
\Large

\refstepcounter{orange}
{\bf \theorange}.\;
\la{intropara}
{\bf Introduction:} In supergravity \ci{freepro,ferrarabook,west,brandtclaim,westandstelle,Weinberg3}  we   start out with Einstein's theory, but it needs to be  formulated with a vierbein $ e_{\m a}  $ and a spin connection
$w_{\m ab}$, instead of the metric tensor $g_{\m\n}$ and an affine  connection $\G_{\m \n \lam}$.  The vierbein   and  spin connection are needed so that the gravitino $ \Y_{\m \a}$ can be coupled to gravity.  All three of these fields are gauge fields, and their gauge transformations are of the form
\be
\d e_{\m}^{\;\; a} = \pa_{\m} \x^a + \cdots
;\;\d w_{\m}^{\;\; a b} = \pa_{\m} \r^{ab} + \cdots
;\; \d \Y_{\m\a} = \pa_{\m} C_{\a} + \cdots
\ee
where $\x^a,\r^{ab}, C_{\a}$ are the Faddeev-Popov ghosts of general coordinate transformations, local Lorentz rotations, and local supersymmetry transformations, respectively.   We can take all these variations $\d$ to be Grassmann odd.  That trick was invented  by BRS and it facilitates the verification of the closure of the algebra of these variations. So $ e_{\m a}, w_{\m ab},C_{\a}$  are all Grassmann even and $ \Y_{\m \a},\x^a ,\r^{ab} $ are all Grassmann odd.

\refstepcounter{orange}
{\bf \theorange}.\;
\la{sugraaction}
{\bf   Supergravity:}  We will not need the actual details here of the supergravity theory or even its action $\cA_{\rm Sugra}$. But it is useful to write down  the action:
\be
\cA_{\rm Sugra} =
\int d^4 x \lt \{ e R[ w , e] + \ve_{\m\n\lam \s} \ovY^{\m} \s^{\n} D^{\lam}[w] \Y^{\s} - \fr{1}{3} B^{\m} B_{\m}  +\fr{1}{3} M \oM 
\rt \}
\la{sugraaction1}
\ee
Here $R[ w,e]$ is Einstein's curvature scalar as a function
of the spin connection and the vierbein,  $e$ is the determinant of the vierbein, and $D_{\m}[w]$ is a covariant derivative formed using the spin connection. It includes two auxiliary fields, $ B^{\m} $ and $M$, which enable the closure of  the variation of the gravitino.

\refstepcounter{orange}
{\bf \theorange}.\;
\la{order}
{\bf   First, second and 1.5  order formalisms for supergravity:} 
In supergravity, it appears that one could choose to use the first order formalism, the second order formalism or the 1.5 order formalism.  They differ by their treatment of the spin connection field $w_{\m ab}$. This field $w_{\m ab}$ seems to be an auxiliary in the theory, and so it can be removed in various ways, or left in place. The only formalism that is actually used at present is the 1.5 order formalism, because it is the simplest.  However, as is shown here, it is also fundamentally flawed.

\refstepcounter{orange}
{\bf \theorange}.\;
\la{organization} {\bf Organization of this Paper:} This paper will discuss all three formalisms for supergravity in a general way, and then illustrate them in detail with the WZ model.

\refstepcounter{orange}
{\bf \theorange}.\; 
\la{puzzleaboutdeltaw}
 {\bf Results of this paper:} It appears that, in the 1.5 order formalism, the form of all the transformations are known except
 $\D_{1}  w_{\m ab}$.  It also appears that the transformation  $\D_{1}  w_{\m ab}$ is not needed. However, as is shown in paragraph \ref{wrongway} below, the 1.5 formalism does not really generate a closed algebra. To corrrect this problem it is necessary to start from the first order formalism, and to find a new nilpotent transformation $\D_{1}  w_{\m ab}$ for the 
spin connection.  But that is of course not the whole story,  since that  closure requires new auxiliary fields, and the transformations of the other fields and the action can be expected to change, to some extent, in order to accomodate the new transformations.
The conclusion of this paper is that
\ben
\item  We do not yet know  the transformation $\D_1 w_{\m ab}$ but there is an obvious guess to make (see paragraph \ref{spinconisYM}), and
\item
 We need the transformation $\D_1 w_{\m ab}$ to formulate any  BRS identity for supergravity. 
\een
There is  a current candidate for a transformation $\D_{1.5}  w_{\m ab}$ in the 1.5 order formulation, as we will describe below.  But we will show that this transformation $\D_{1.5}  w_{\m ab}$  has an incurable problem, and, as a result,it follows that we do not yet have a correct formulation for  supergravity in 3+1 dimensions.

\refstepcounter{orange}
{\bf \theorange}.\;
\la{order1}
{\bf   First  Order Formalism for supergravity:} In the first order formalism, the vierbein, the spin connection and the gravitino are all considered to be fundamental variables, and the equations of motion are obtained by varying each of them independently. It is called first order because it turns out that the equations of motion here  for the fields $e_{\m a}$ and $w_{\m ab}$  are of first order in the derivatives.

\refstepcounter{orange}
{\bf \theorange}.\;
\la{transofgrav}
{\bf The first order  identity for supergravity} might be expected to be of the form\be
\cM_{\rm Sugra}= \int d^4 x \lt \{
\fr{\d \cA_{\rm Sugra}}{\d e_{\m a}} \D_1 e_{\m a}  
+
\fr{\d \cA_{\rm Sugra}}{\d \Y_{\m } }\D_1 \Y_{\m }  
+
\fr{\d \cA_{\rm Sugra}}{\d \Y_{\m } }\D_1 \ovY_{\m }  
\ebp
+
\fr{\d \cA_{\rm Sugra}}{\d M }\D_1 M
+
\fr{\d \cA_{\rm Sugra}}{\d \oM }\D_1 \oM
+
\fr{\d \cA_{\rm Sugra}}{\d B^{\m} }\D_1 B^{\m} 
+
\fr{\d \cA_{\rm Sugra}}{\d w_{\m ab} } \D_1  w_{\m ab}
\rt \}=0
\la{sugraformula}
\ee
This identity would summarize all the invariances of the action. 
The variations $\D e_{\m a}, \D \Y_{\m},\cdots$ are all complicated local functions of the variables $e_{\m a}, \Y_{\m},\cdots$ and they also are functions of the three ghosts of supergravity.  These three ghosts $\x^a,\r^{ab}, C_{\a}$ were mentioned in paragraph
 \ref{intropara} above. 
To close the algebra we also need the variations of the ghosts 
$\D \x^a,\D \r^{ab}, \D C_{\a}$, which are also functions of all the variables, including the ghosts themselves. 

\refstepcounter{orange}
{\bf \theorange}.\; 
\la{deltonesquared}
{\bf Closure of the algebra for the first order  identity for supergravity} would mean that  the  variations of the variations are all zero. 
\be
\D_1^2 e_{\m a} =0, \D_1^2 \Y_{\m} =0, \D_1^2 \ovY_{\m} =0, 
 \eb
\D_1^2 B_{\m } =0,  \D_1^2 M =0, \D_1^2 \oM =0, \D_1^2 w_{\m a b} =0, \D_1^2 \r_{ a b} =0,  \eb
\D_1^2 \x_{\m } =0,  \D_1^2 C_{ \a}=0,  \D_1^2 \oC_{\dot \a} =0
\la{firstorderclosure}
\ee
Again the operator $\D_1$ here is Grassmann odd. These transformations are the starting point for the BRS cohomology   \ci{Weinberg2, taylor,poissonbrak,  {Becchi:1975nq}, {zinnbook},{Zinnarticle},{becchiarticles1},
becchiarticles2}  of supergravity.  In this paper,  we will not need the explicit forms of these variations and we will not contemplate  any other transformations, such as those suggested in paragraph \ref{spinconisYM}.

 \refstepcounter{orange}
{\bf \theorange}.\; 
\la{puzzle2} {\bf Early in the history of  supergravity,} the paper \ci{deserandzumino} did contain some pieces of $\D_1 w_{\m ab}$.  But that was before the auxiliary fields $B_{\m}$ and $M$ had been discovered.  Since then it appears that there has been little discussion of $\D_1 w_{\m ab}$, because the 1.5 order formalism has been used. 

\refstepcounter{orange}
{\bf \theorange}.\;
\la{transofgrav2}
{\bf In the second order formalism,} the idea is to   replace the spin connection $w_{\m ab}$ with its expression as a function of the vierbein and the gravitino, obtained by finding the equation of motion of the spin connection:
\be
\fr{\d \cA_{\rm Supergravity}}{\d w_{\m ab} }=0.
\la{eqmtforw}
\ee
Then the theory would be expressed in terms of the vierbein and the gravitino, and one would proceed to deal with  the equations of motion obtained by varying each of those independently.  The spin connection would play no further role in the formalism. This is called second order because it turns out that the equations of motion for the field $e_{\m a}$ here are of second order in the derivatives. 
For the second order formalism, we would   expect to find an identity of the form
\be
\cM_{\rm Second}= \int d^4 x \lt \{
\fr{\d \cA_{\rm Second}}{\d e_{\m a}} \D_2 e_{\m a}  
+
\fr{\d \cA_{\rm Second}}{\d \Y_{\m } }\D_2 \Y_{\m }  
+
\fr{\d \cA_{\rm Second }}{\d \Y_{\m } }\D_2 \ovY_{\m }  
\ebp
+
\fr{\d \cA_{\rm Second }}{\d M }\D_2 M
+
\fr{\d \cA_{\rm Second }}{\d \oM }\D_2 \oM
+
\fr{\d \cA_{\rm Second}}{\d B^{\m} }\D_2 B^{\m} 
\rt \}=0
\la{secondformula}
\ee
Note that the term $\fr{\d \cA_{\rm Sugra}}{\d w_{\m ab} } \D_1  w_{\m ab}
$, which was present at the end of (\ref{sugraformula}) has been removed from  (\ref{secondformula}). There is no term of the form $\fr{\d \cA_{\rm Sugra}}{\d w_{\m ab} } \D_2  w_{\m ab}
.$  This identity would summarize all the invariances of the action. The action here would be obtained from the action 

\be
\cA_{\rm Second}
=\cA_{\rm Sugra}[ e, \Y, \ov \Y, w \ra w[e, \Y, \ovY]]
\ee
where the expression  $w[e, \Y, \ovY]$ is obtained from the solution of 
\be
\fr{\d \cA_{\rm Sugra}}{\d w_{\m ab} } =0,
\ee
which is linear in $w$.

\refstepcounter{orange}
{\bf \theorange}.\; 
\la{closealgsugra2}
{\bf By closure of the algebra here, for the second order formalism} we again mean that  the  variations of the variations are all zero:
\be
\D_2^2 e_{\m a} =0, \D_2^2 \Y_{\m} =0, \D_2^2 \ovY_{\m} =0, 
\D_2^2 B_{\m } =0,  \D_2^2 M =0,  \ee
\be  \D_2^2 \oM =0,\D_2^2 \r_{ a b} =0,  \D_2^2 \x_{\m } =0,  \D_2^2 C_{ \a}=0,  \D_2^2 \oC_{\dot \a} =0
\la{secondorderclosure}
\ee
 However the variations $\D_2 $ here are clearly not the same as the $\D_1$  were for the identity 
(\ref{sugraformula}), because the field $w_{\m ab}$ has been eliminated, and so it has no variation. We shall return to the question of the construction of these new variations 
$\D_2 $ below.   We will show  that if we had   variations satisfying the equations in paragraph \ref{firstorderclosure}, then we could construct solutions for 
$\D_2$ which satisfy $\D_2^2=0$.  However we would need to expand the equations  (\ref{secondorderclosure}) to include some sources. To do that construction we would need to introduce the BRS formalism in the form of the Master Equation.

\refstepcounter{orange}
{\bf \theorange}.\;
\la{missingtransgrav1.5}
{\bf   The 1.5 order formalism tries to take advantage} of  the merits of both the first order formalism, and the second order formalism. 
The action is expressed as a function of the same variables as in the first order formulation (because  that is the simplest formulation).  All three variables are varied, as in the first order formalism.  But then there is an effort to implement the equation of motion of the spin connection so that its independent variation does not complicate things.  

\be
\cM_{\rm Sugra\;With\;1.5\;Trick}= \int d^4 x \lt \{
\fr{\d \cA}{\d e_{\m a}} \D_{1.5} e_{\m a}  
+
\fr{\d \cA}{\d \Y_{\m } }\D_{1.5} \Y_{\m }  
+
\fr{\d \cA}{\d \Y_{\m } }\D_{1.5} \ovY_{\m }  
\ebp+
\fr{\d \cA}{\d M }\D_{1.5} M +
\fr{\d \cA}{\d \oM }\D_{1.5} \oM 
+
\fr{\d \cA}{\d B^{\m} }\D_{1.5} B^{\m} 
\rt \}=0
\ee

The other attraction of the 1.5 order formalism is that it looks like it avoids the problem of finding a form for  $\D_{1} w_{\m ab}$.   
But in fact it does not succeed in that avoidance, as will be shown.

\refstepcounter{orange}
{\bf \theorange}.\;
\la{closureof1.5}
{\bf   We would also expect} to have the same form as for the second order formalism in (\ref{secondformula}):

\be
\D_{1.5}^2 e_{\m a} =0, \D_{1.5}^2 \Y_{\m} =0, \D_2^2 \ovY_{\m} =0, 
\D_{1.5}^2 B_{\m } =0,  \D_{1.5}^2 M =0, \eb
\D_{1.5}^2 \oM =0, 
 \D_{1.5}^2 \r_{ a b} =0,  \D_{1.5}^2 \x_{\m } =0,  \D_{1.5}^2 C_{ \a}=0,  \D_{1.5}^2 \oC_{\dot \a} =0
\la{1.5orderclosure}
\ee

We might ask at this point whether we expect these variations $\D_{1.5}$
 to be the same as 
$\D_{1}$
or $\D_{2}$ or different from both of them, and how we can find them.   In fact this is not a real question, because the 1.5 formalism is not a valid formulation.  An impatient reader here can look at paragraph \ref{wrongway} to see how this happens.

\refstepcounter{orange}
{\bf \theorange}.\;
\la{problem}
{\bf The easy way to find $\D_{1.5} w_{\m ab}$:} It has been suggested in the literature that we can find the variation $\D_{1.5} w_{\m ab}$ by using the equation of motion (\ref{eqmtforw}) to solve for  $ w_{\m ab}$ as a function of $e_{\m a}$ and $\Y_{\m}$, and then construct $ \D_{1.5} w_{\m ab}$ by using $\D_1 e_{\m a}$ and $\D_1 \Y_{\m}$ in that expression to write down  $\D_{1.5} w_{\m ab}$.

\refstepcounter{orange}
{\bf \theorange}.\;
\la{problem2}
{\bf This easy route to find $\D_{1.5} w_{\m ab}$ does not work.}  We do not need to get into the terrific complexity of supergravity to show this. We can explain the problem using the first supersymmetric action that was discovered: the Wess-Zumino model. In that way we will see that if we use the equation of motion to write everything in terms of the fields other than $w_{\m ab}$, then the nilpotence fails.  As mentioned above, an impatient reader here can look at paragraph \ref{wrongway} to see how this happens.

\refstepcounter{orange}
{\bf \theorange}.\;
\la{intropar4}
{\bf The three kinds of order formalisms in the WZ multiplet:} We can easily analyze the basic features of the three kinds of   order formalism in a very much simpler model. Instead of dealing with the spin connection in supergravity we will deal with the auxiliary fields $F, \oF$ in the WZ model. The same idea applies to any auxiliary field in any theory. For the WZ multiplet, we will start with the first order, then generate the second order, and then contemplate the 1.5 order.
Then we will return to the situation  in supergravity.

 \refstepcounter{orange}
{\bf \theorange}.\;
\la{intropar5}
{\bf  Here is the action for the WZ model,} including a simple mass term of the chiral type. We use two component complex spinors, because the theory is naturally adapted to them.
\be
\cA_{\rm Fields} = \int d^4 x \lt \{
\pa_{\m} A \pa^{\m} \A
+ \c^{\a}  \s^{\m\dot \b}_{\a} \pa_{\m} {\ov \c}_{\dot \b}
+ F \oF 
\ebp
+ m \lt ( A F - \fr{1}{2}\c^{\a}\c_{\a} \rt ) 
+ m \lt ( \A \oF - \fr{1}{2} {\ov \c}^{\dot \a}{\ov \c}_{\dot \a} \rt ) 
\rt \}
\ee
This action obeys an identity that arises from supersymmetry.  We define the following variations:
\be
\D_{1} A = C^{\a} \c_{\a}
;\; \D_{1} \c_{\a} =  \s^{\m\dot \b}_{\a}\pa_{\m} A \oC_{\dot \b}
+ F C_{\a};\;
\D_{1} F = \oC_{\dot \b}  \ov \s^{\m\dot \b}_{\a}\pa_{\m} \c^{\a}
\la{deltaF}\ee
\be
\D_{1} \A = \oC^{\dot \a} {\ov \c}_{\dot \a}
;\; 
\D_{1} {\ov \c}_{\dot\a} =  \ov\s^{\m \b}_{\dot\a}\pa_{\m} \A C_{\b}
+ \oF C_{\dot\a}
;\;
\D_{1}\oF = C_{\b}  \s^{\m \b}_{\dot\a}\pa_{\m} {\ov \c}^{\dot \a}
\la{deltaFCC}\ee
This is rigid supersymmetry, so we assume that the ghosts $  C_{\a}, \oC_{\dot\a}$ are constants, independent of the coordinates, so that
\be
\D_{1} C_{\a} = 
\D_{1} \oC_{\dot\a} =0;\;
\pa_{\m}   C_{\a} =\pa_{\m} \oC_{\dot\a}=0
\ee

\refstepcounter{orange}
{\bf \theorange}.\;
\la{firstorderidentityforWZ}
{\bf The First Order Identity:}  The following is the identity in this theory that is generated by its supersymmetry:
\[
\cM_{\rm WZ}= \int d^4 x \lt \{
\fr{\d \cA}{\d A} \D_{1} A+
\fr{\d \cA}{\d \c_{\a}} \D_{1} \c_{\a}
+ \fr{\d \cA}{\d F} \D_{1} F
\rt.
\]
\be
\lt.
\fr{\d \cA}{\d \A} \D_{1} \A+
\fr{\d \cA}{\d {\ov \c}_{\dot\a}} \D_{1} {\ov \c}_{\dot\a}+
\fr{\d \cA}{\d \oF} \D_{1} \oF
\rt \}=0
\la{includeF}
\ee
By saying that this is an identity, we mean that adding all the terms together gives us exactly zero, without imposing any equation that is assumed to be satisfied by the fields. Formula (\ref{includeF})  is simply true, no matter what the values of the fields are.  This can be done explicitly in one page, and it looks very much like paragraph \ref{explicitinvarianceofaction} below.

\refstepcounter{orange}
{\bf \theorange}.\;
\la{nilpotence}
{\bf   A closely related identity is the nilpotence identity:}
 Thus for example we have
\be
\D_{1}^2 A =
 C^{\a} \D_{1} \c_{\a}= 
( C^{\b}  \s^{\m\dot \g}_{\b}\oC_{\dot \g} )\pa_{\m} A
;\;
\D_{1}^2 \c_{\a} =( C^{\b}  \s^{\m\dot \g}_{\b}\oC_{\dot \g} )\pa_{\m} \c_{\a} 
\la{goodresult1chi}
\ee
Similar results hold for all the fields in the formulae above.
The operator $\D_{1}^2$ gives rise to a derivative times the constant $ C^{\a}  \s^{\m\dot \b}_{\a}\oC_{\dot \b}$. This can be treated as zero, because if we take the derivative of the Lagrangian, we get zero variation for the action, because the integral of a total derivative is taken to be zero in quantum field theory.

\refstepcounter{orange}
{\bf \theorange}.\;
{\bf   The second order formalism for the WZ model:} 
To generate this, we can start with the first order formalism above.  The idea here is to first formulate the BRS ZJ Master Equation for the first order formalism.
Then we will integrate the auxiliary fields $F,\oF$ out of the theory.  That will lead us to a Master Equation that does not refer to the auxiliary fields at all.

\refstepcounter{orange}
{\bf \theorange}.\;
{\bf    Formulating the Master Equation for the WZ model:} 
The action we need to use is 
\be
\cA_{\rm Total}= 
\cA_{\rm Fields}+
\cA_{\rm ZJ\;Sources}
\la{totalactionforWZ}
\ee
where we define the ZJ Source Action $\cA_{\rm ZJ\;Sources}$:
\be
\cA_{\rm ZJ\;Sources} = \int d^4 x \lt \{
\widetilde{A}\D_{1} A+
\widetilde{\A}\D_{1} \A+
\widetilde{\c}_{\a} \D_{1}  \c^{\a}
\ebp
+
\widetilde{{\ov \c}_{\dot \a} }\D_{1} {\ov \c}^{\dot \a} +
\widetilde{F}\D_{1} F+
\widetilde{F}\D_{1} \oF
\rt \}
\ee

Let us write this explicitly using the above expressions in ({refdeltaF}) through (\ref{deltaFCC}):
\be
\cA_{\rm ZJ\;Sources} = \int d^4 x \lt \{
\widetilde{A}C^{\a} \c_{\a}+
\widetilde{\A}\oC^{\dot \a} {\ov \c}_{\dot \a}+
\widetilde{\c}^{\a} \lt (\s^{\m\dot \b}_{\a}\pa_{\m} A \oC_{\dot \b}
+ F C_{\a}\rt )
\ebp+
\widetilde{{\ov \c}^{\dot \a} }\lt (  \ov\s^{\m \b}_{\dot\a}\pa_{\m} \A C_{\b}
+ \oF C_{\dot\a}\rt )
+
\widetilde{F}\oC_{\dot \b}   \s^{\m\dot \b}_{\a}\pa_{\m} \c^{\a}+
\widetilde{F} C_{\b}  \ov\s^{\m \b}_{\dot\a}\pa_{\m} {\ov \c}^{\dot \a}
\rt \}
\ee
  This will be the analogue of paragraph \ref{transofgrav}, but for this much simpler theory. 

\refstepcounter{orange}
{\bf \theorange}.\;
{\bf    Formulating the Master Equation in General for the First Order Formalism:} This is a standard technique in BRS theories, but we want to emphasize the points that are crucial to show that the 1.5 order formalism does not work. It is easiest to formulate this in a general way.  We suppose that there is a set of fields $\f_i$ which may be physical fields, auxiliaries, ghosts or antighosts, and which may have even or odd Grassmann character. We suppose that there is a local action 
\be
\cA[\f]= \int d^4 x \cL[\f]
\ee
and a set of transformations $\D_1 \f^i$ which satisfy the equations
\be
\sum_i \int d^4 x \;\fr{\d \cA[\f]}{\d \f^i}\D_1 \f^i = 0
;\;\D_1^2 \f^i =0
\la{phitransinv}
\ee

We define the action by
\be
\cA_{\rm Integrand} = \cA[\f,\widetilde{\f}]
+ \cA[{\rm Field\;Sources}]
;\;
cA[\f,\widetilde{\f}] = \cA[\f]
+ \cA[{\rm ZJ\;Sources}]
\ee
\be
\cA[{\rm ZJ\;Sources}]
= \sum_i \int d^4 x \widetilde{\f}_i \D_1 \f^i
;\;\cA[{\rm Field\;Sources}]= \sum_i \int d^4 x \widehat{\f}_i \f^i
\ee

\refstepcounter{orange}
{\bf \theorange}.\;
{\bf 
The Master Equation:}
Now we consider the functional path integral:
\be
Z[  \widetilde{\f}_i ,\widehat{\f}_i]
= \P_{\rm i, x} \int d \; {\rm \f^i}(x) \;\;
e^{\fr{i}{\hbar} \cA_{\rm Integrand}}
\la{phipathintegral}
\ee
Then we see that there is an identity
\be
\sum_i \int d^4 x \;  \widehat{\f}_i \fr{\d Z}{\d \widetilde{\f}_i } = \sum_i \int d^4 x \;  \widehat{\f}_i \D_1 \f^i = 0
\la{theidentityforphi}
\ee
This arises from a transformation 
\be
\f^i \ra \f^i + \e \D_1 \f^i 
\ee 
on  the integration variables in (\ref{phipathintegral})
 and using the equations 
(\ref{phitransinv}) ($\e$ is a Grassmann odd parameter to make the Grassmann characters match).  Now we define the 1PI generating functional $\cG_{\rm Connected}$ which generates connected diagrams by
\be
Z = e^{\fr{i}{\hbar}\cG_{\rm Connected}}
\ee
and then we define the  1PI vertex  generating functional  $\cG[ \f]$, which is a function of  new classical variables $\f^i$, by the Legendre transform
\be
\cG[\widehat{\f}] = \cG[ \f] -\sum_i 
 \int d^4 x \f^i  \widehat{\f}_i  
\ee
It follows that
\be
\widehat{\f}_i = \fr{\d \cG[\f]}{\d  \f^i }
\ee

Now we can rewrite the identity (\ref{theidentityforphi}) in the form 
\be
\cM=     \sum_i^N
\int d^4 x \; \fr{\d \cG }{\d {\rm \f}^i} \fr{\d \cG}{\d \widetilde{\f}_i}  = 0
\la{1piequation}
\ee
Here we are supposing that there are $N$ fields.
We can expand in the parameter $\hbar$ and the first term is the action
\be
 \cG = cA[\f,\widetilde{\f}]  + O(\hbar)
\ee
So we should find that 
\be
\cM[ \cA[\f,\widetilde{\f}]]=  \sum_i^N \int d^4 x \; \fr{\d  \cA[\f,\widetilde{\f}]   }{\d {\rm \f}^i} \fr{\d  \cA[\f,\widetilde{\f}]  }{\d \widetilde{\f}_i}  = 0
\la{eqforBRSaction}
\ee
 The identity (\ref{eqforBRSaction}) is the same as (\ref{includeF}), for the case of the WZ multiplet.

\refstepcounter{orange}
{\bf \theorange}.\;
{\bf The consequence of the above is to show} that for any theory, including supergravity, in order to formulate the BRS Master Equation for that theory, we need to have the theory satisfy two requirements:
\ben
\item
The action must be invariant.
\item
The variations must be nilpotent.
\een
We have stated these requirements  for the WZ model in the first order formalism in paragraphs \ref{firstorderidentityforWZ} and \ref{nilpotence}. Also, we have assumed these requirements, in the first order formalism, in paragraphs \ref{transofgrav}
and \ref{deltonesquared}, for supergravity.

\refstepcounter{orange}
{\bf \theorange}.\;
{\bf We will now show how to derive the second order formalism from the first order formalism for the WZ model:} This uses the BRS Master Equation in a fundamental way. We expect all theories and all integrations of all auxiliary fields to behave in the same way, so in particular this should be an exact analogue of what is needed to complete the second order formalism for supergravity discussed in paragraphs \ref{transofgrav2}
and \ref{closealgsugra2}. 

\refstepcounter{orange}
{\bf \theorange}.\;
{\bf How to Integrate an Auxiliary out of the Action:}
All we need to do to accomplish this is to drop the ZJ sources and the Field Sources for the auxiliary, and then to perform the integration of the auxiliary explicitly in the path integral.  This amounts to doing the path integral:
\be
Z
= \P_{\rm x} \int d F(x)  d  \oF(x) \;\; 
e^{\fr{i}{\hbar} \int d^4 x \;\ \lt \{ -   F + \om \A 
 + \widetilde{\c}^{\a}   C_{\a}  \rt \} \lt \{-  \oF + m A + \widetilde{{\ov \c}}^{\dot \a}   \oC_{\dot \a}
\rt \}
}
\eb
e^{-\fr{i}{\hbar} \int d^4 x \;\ \lt \{  \om \A 
 + \widetilde{\c}^{\a}   C_{\a}  \rt \} \lt \{  m A + \widetilde{{\ov \c}}^{\dot \a}   \oC_{\dot \a}
\rt \}
}
= 
e^{-\fr{i}{\hbar} \int d^4 x \;\lt \{  \om \A 
 + \widetilde{\c}^{\a}   C_{\a}  \rt \} \lt \{  m A + \widetilde{{\ov \c}}^{\dot \a}   \oC_{\dot \a}
\rt \}
}
\la{integrateF}
\ee
where we shift $F$ in the first term to do the integral, and we absorb an  irrelevant constant. All the terms in the action that contain the auxiliaries $F$ or $\oF$ are dropped from the action at the same time.

\refstepcounter{orange}
{\bf \theorange}.\;
{\bf Starting with the action \ref{totalactionforWZ}, the  total equations of motion of F}, used above,  are 
\be
 \fr{\d \cA_{\rm Total}}{\d F^i}
=
-  \oF + m A + \widetilde{{\ov \c}}^{\dot \a}   \oC_{\dot \a}
;\; \fr{\d \cA_{\rm Total}}{\d \oF}
=
-   F + \om \A 
 + \widetilde{\c}^{\a}   C_{\a}
\ee

Let us suppose that we put the action into a path integral and integrate the fields $F,\oF$.  We get the new action

\be
\cA_{\rm New}= \cA_{\rm  Total\;with\; F\ra 0, \widetilde{F} \ra 0,  \widehat{F} \ra 0}
+ \cA_{\rm From \; integration\; of\; F} 
\ee
where\be
\cA_{\rm  Total\;with\; F\ra 0, \widetilde{F} \ra 0,  \widehat{F} \ra 0} 
\eb= \int d^4 x \lt \{
\pa_{\m} A \pa_{\m} \A
+ \c^{\a}  \s^{\m\dot \b}_{\a} {\ov \c}_{\dot \b}
  + m \lt ( - \c^{\a}\c_{\a} \rt ) 
+ m \lt (  - {\ov \c}^{\dot \a}{\ov \c}_{\dot \a} \rt ) 
\rt \}
\eb
+
 \int d^4 x  \lt \{
\widetilde{A}C^{\a} \c_{\a}+
\widetilde{\A}\oC^{\dot \a} {\ov \c}_{\dot \a}+
\widetilde{\c}^{\a} \lt [ \s^{\m\dot \b}_{\a}\pa_{\m} A \oC_{\dot \b} \rt ]
+\widetilde{{\ov \c}^{\dot \a} }
\lt [ \ov\s^{\m \b}_{\dot\a}\pa_{\m} \A C_{\b} \rt ]
\rt \}
\ee
\be
 \cA_{\rm From \; integration\; of\; F} = 
- \int d^4 x  \lt \{
 m A  + \widetilde{\c}^{\a}   C_{\a}
\rt \}
\lt \{  m \A  + \widetilde{{\ov \c}}^{\dot \a}   \oC_{\dot \a}
\rt \}
\ee
and we expect the new {identity}:

\be
\cM_{\rm New}= \int d^4 x \lt \{
\fr{\d \cA_{\rm New }}{\d A} 
\fr{\d \cA_{\rm New}}{\d \widetilde{A}} 
+
\fr{\d \cA_{\rm New }}{\d \c_{\a}}
\fr{\d \cA_{\rm New}}{\d \widetilde{\c}^{\a}}
\ebp+
\fr{\d \cA_{\rm New }}{\d \cA_{\rm New }} 
\fr{\d \cA_{\rm New}}{\d \widetilde{\A}} 
+
\fr{\d \cA_{\rm New }}{\d {\ov \c}_{\dot\a}}  
 \fr{\d \cA_{\rm New}}{\d \widetilde{{\ov \c}}^{\dot\a}}
\rt \} =0
\la{reducedidenttiyfornew}
\ee

\refstepcounter{orange}
{\bf \theorange}.\;
\la{explicitinvarianceofaction}
{\bf    Let us verify that this is an identity, and observe how it works:}
We have 
\be
\int d^4 x \fr{\d \cA_{\rm New }}{\d A} 
\fr{\d \cA_{\rm New}}{\d \widetilde{A}} 
=
\int d^4 x  \lt \{
- \Box \A-\pa_{\m} \widetilde{\c}_{\a}  \s^{\m\dot \b}_{\a}   \oC_{\dot \b} - 
 m
\lt (  \om \A  + \widetilde{{\ov \c}}^{\dot \a}   \oC_{\dot \a}
\rt )
 \rt \}
 \lt \{  
C^{\a} \c_{\a}\rt \}
\ee
\be
\int d^4 x \fr{\d \cA_{\rm New }}{\d \c_{\a}}
\fr{\d \cA_{\rm New}}{\d \widetilde{\c}^{\a}}
= \int d^4 x \lt \{     \s^{\n \a\dot \g} \pa_{\n} {\ov \c}_{\dot \g}
  + m    \c^{\a}  
 \rt \}
 \lt \{ \s^{\m\dot \b}_{\a}\pa_{\m} A \oC_{\dot \b} 
+
  \om \A C_{\a}  + \widetilde{{\ov \c}}^{\dot \a}   \oC_{\dot \a}C_{\a}
\rt \}
\ee
\be
\int d^4 x \fr{\d \cA_{\rm New }}{\d \A} 
\fr{\d \cA_{\rm New}}{\d \widetilde{\A}} 
=
\int d^4 x  \lt \{
- \Box A-\pa_{\m} \widetilde{{\ov \c}}_{\dot\a}  \ov\s^{\m\b}_{\dot \a}   C_{\b} - 
 \om
\lt (  m A  + \widetilde{\c}^{\a}   C_{\a}
\rt )
 \rt \}
 \lt \{  
\oC^{\dot\a} {\ov \c}_{\dot\a}\rt \}
\ee 
\be
\int d^4 x \fr{\d \cA_{\rm New }}{\d {\ov \c}_{\dot\a}}
\fr{\d \cA_{\rm New}}{\d \widetilde{{\ov \c}}^{\dot\a}}
= \int d^4 x \lt \{    \ov \s^{\n \dot\a  \g} \pa_{\n} \c_{\g}
  + \om    {\ov \c}^{\dot \a}  
 \rt \}
 \lt \{ \ov\s^{\m\b}_{\dot\a}\pa_{\m} \A C_{\b} 
+
  m A \oC_{\dot\a}  + \widetilde{\c}^{ \a} C_{\a}  \oC_{\dot \a}
\rt \}
\ee
It is a simple matter to pick out the cancelling pairs in the above set of objects. One gets a total of zero, thus verifying (\ref{reducedidenttiyfornew}). 

\refstepcounter{orange}
{\bf \theorange}.\;
{\bf   Also we now have nilpotence.} But the nilpotence involves both the variations of the fields and the variation of the ZJ sources.
Thus   we have, for example, 
\be
\D  {\c}_{\a}= \fr{\d \cA_{\rm New}}{\d \widetilde{\c}^{\a}}
= 
 \lt \{ \s^{\m\dot \b}_{\a}\pa_{\m} A \oC_{\dot \b} 
+
  \om \A C_{\a}  + \widetilde{{\ov \c}}^{\dot \a}   \oC_{\dot \a}C_{\a}
\rt \}
\ee
and
\be
\D \widetilde{A}=  \fr{\d \cA_{\rm New }}{\d A} 
= \lt [
\Box \A-\pa_{\m} \widetilde{\c}_{\a}  \s^{\m\dot \b}_{\a}   \oC_{\dot \b} - 
 m
\lt (  \om \A  + \widetilde{{\ov \c}}^{\dot \a}   \oC_{\dot \a}
\rt )
 \rt ]
\ee
and we can derive nilpotence equations quite easily for all the fields and all the sources.  For example we can show that:
\be
\D^2 \c_{\a}= 
  \oC_{\dot \b}   \s^{\m\dot \b}_{\a} C^{\a}\pa_{\m}  \c_{\a}
;\;
\D^2 \widetilde{A}= 
  \oC_{\dot \b}   \s^{\m\dot \b}_{\a} C^{\a}\pa_{\m}  \widetilde{A}
\la{correct2orderforWZ}
\ee

\refstepcounter{orange}
{\bf \theorange}.\;
{\bf  So, using the ZJ sources, and the Master equation for the first order formalism,} we have derived the second order formalism by integrating the auxiliary F using the path integral.

\refstepcounter{orange}
{\bf \theorange}.\;
{\bf  Using the path integral, one can derive the BRS identity quite easily in the form (\ref{reducedidenttiyfornew})
 for the one particle irreducible vertices.} But this requires that we start with invariance and nilpotence for all the fields, including the ones we are going to integrate out of the theory ($F,\oF$ in this case, and the three auxiliaries  $M, B_{\m},w_{\m a b}$  for supergravity).

\refstepcounter{orange}
{\bf \theorange}.\;
\la{intropar3}
{\bf   The following is the 1.5 order formalism for the WZ theory:}  It is not an identity. The following  is exactly the same as what is being done in supergravity with the 1.5 trick as described in paragraph \ref{missingtransgrav1.5}.  
\be
\cM_{\rm WZ\;With\;1.5\;Trick}= \int d^4 x \lt \{
\fr{\d \cA}{\d A} \D A+
\fr{\d \cA}{\d \A} \D \A+
\fr{\d \cA}{\d \c_{\a}} \D \c_{\a}
+
\fr{\d \cA}{\d {\ov \c}_{\dot\a}} \D {\ov \c}_{\dot\a}
\rt \}=0
\la{leaveoutF}
\ee
If we evaluate the expression (\ref{leaveoutF}), we do not get zero.  We can say that (\ref{leaveoutF}) is true, if, and only if, we use the equation of motion of F and $\oF$.  
The missing tems are
\be
\int d^4 x \fr{\d \cA_{\rm New }}{\d F}
\D F
= \int d^4 x \lt \{ \oF + m A  
\rt \} \lt \{ \oC_{\dot \b}   \s^{\m\dot \b}_{\a}\pa_{\m} \c^{\a}\rt \}
\ee
\be
\int d^4 x \fr{\d \cA_{\rm New }}{\d \oF}
\D \oF
= \int d^4 x \lt \{  F + m \A  
\rt \}
\lt \{ C_{\b}  \ov\s^{\m \b}_{\dot\a}\pa_{\m} {\ov \c}^{\dot \a}
\rt \}
\ee
and, indeed, they vanish by the equations of motion of $F$ and $\oF$.

\refstepcounter{orange}
{\bf \theorange}.\;
{\bf
The value of the BRS formalism} occurs when there is an identity, not when there is an identity which also requires the  use of  the equations of motion in addition.  If we are content to use the equations of motion we could actually say that each individual term is zero:
\be
\fr{\d \cA}{\d A}=
\fr{\d \cA}{\d \A} =
\fr{\d \cA}{\d \c_{\a}} =
\fr{\d \cA}{\d {\ov \c}_{\dot\a}}
= \fr{\d \cA}{\d F}  
=\fr{\d \cA}{\d \oF}  
\ee
In that case we would not need any of the $\D$ variations at all.  Now one might argue that  $F$ and $\oF$ are auxiliary fields, and so they are special, so that nothing goes wrong if we use the 1.5 order formalism for auxiliary fields.  But that is not true. Something very serious does go wrong, as we now show.

\refstepcounter{orange}
{\bf \theorange}.\;
\la{wrongway}
{\bf  The Wrong Way to Find $\D w_{\m ab}$:}
As mentioned above in paragraph \ref{problem}, sometimes we are told to find $\D w_{\m ab}$ by applying the variations of the other fields to its form as a function of the other fields, found by using the equation of motion. The claim here is that that idea does not work.   To show it does not work, let us try the same idea here, for the simple case of the WZ model, except that here we are going to use the idea to generate $\d F$:
\ben
\item
If one adds all the   terms in (\ref{includeF}) or in (\ref{reducedidenttiyfornew}), one gets identically zero--there is no need to assume an equation of motion.  This was shown explicitly in detail in paragraph \ref{explicitinvarianceofaction}. 
\item
This is not true for the expression 
(\ref{leaveoutF}).  In fact   the equations that need to be used there, to get zero in  (\ref{leaveoutF}), are:
\be
 \fr{\d \cA}{\d F} =- \oF + mA =0;\;
 \fr{\d \cA}{\d F} =- F + m\A =0
\ee
\item
Using the philosophy of the 1.5 formalism, we need to assume that the equation of motion of the auxiliary can be used to prove  (\ref{leaveoutF}), and this means here that 
\be
F = m\A 
\ee
\item
As was mentioned above, we are told in the literature that we can take this equation to get the variation of F, by using the known variations of the other fields.  That would mean that we should take 
\be
\D A = C^{\a} \c_{\a}
;\;
\D \c_{\a}
 =\s^{\m\dot \b}_{\a}\pa_{\m} A \oC_{\dot \b}
+ F C_{\a}
 \lra \s^{\m\dot \b}_{\a}\pa_{\m} A \oC_{\dot \b}
+ m \A C_{\a}
\ee
\be
\D F \lra m \D \A = m 
\oC^{\dot \a} {\ov \c}_{\dot \a}
\ee
\item
Is that actually correct?  Now of course the field $F$ does not actually appear in the final result.  But there are two fields (and their complex conjugates) that do appear.  Let us look at the square $\D^2 $ of their variations, to see if the algebra closes:
\ben
\item
From the above we get:
\be
\D^2 A = C^{\a} ( \s^{\m\dot \b}_{\a}\pa_{\m} A \oC_{\dot \b}
+ m \A C_{\a})
= ( C^{\a}  \s^{\m\dot \b}_{\a}\oC_{\dot \b} )\pa_{\m} A 
\ee
  So that has not changed.  This expression is nilpotent. 
\item
However
\be
\D^2 \c_{\a}
=\s^{\m\dot \b}_{\a}\pa_{\m} \D A \oC_{\dot \b}
+ m \D \A C_{\a}
 =\s^{\m\dot \b}_{\a}\pa_{\m} ( C^{\b} \c_{\b})\oC_{\dot \b}
+ m ( \oC^{\dot \a} {\ov \c}_{\dot \a}) C_{\a}
\la{badresult}
\ee
\item
Here we expect, and we need,  to get 
\be
\D^2 \c_{\a} =( C^{\b}  \s^{\m\dot \g}_{\b}\oC_{\dot \g} )\pa_{\m} \c_{\a} 
\la{goodresult}
\ee
\item
Equation (\ref{badresult}) is not the derivative form (\ref{goodresult}). What has gone wrong?   
\item
What is happening of course, is that instead of using the original result for $\D F$, which was
\be
\D F = \oC_{\dot \b}   \s^{\m\dot \b}_{\a}\pa_{\m} \c^{\a}
\la{originaldeltaF}
\ee
we are using the new result for $\D F$
\be
\D F   = m 
\oC^{\dot \a} {\ov \c}_{\dot \a},
\la{newdeltaF}
\ee
 and they are not the same.  So if we find $\D F$ by evaluating the variation of its value using the equation of motion of $F$, we lose the closure of the algebra.  This then spoils the nilpotence
that would be present in (\ref{goodresult}) and yields 
(\ref{badresult}) instead. The correct result   (\ref{goodresult}) requires (\ref{originaldeltaF}) rather than (\ref{newdeltaF}).

\item
There is a loss of information when we abandon the original variation 
in favor of the new variation.  To preserve the closure of the algebra we need to go through the work of getting the proper second order formulation, as we did above to get equation (\ref{correct2orderforWZ}).  And to do that we need to start with a correct formulation of the first order formalism, as we did above to establish the master equation for the WZ model. 
\een
\item  We would find a similar result for supergravity, if we substitute the auxiliaries $F \ra w_{\m ab}$ appropriately. 
\item
So we can expect to find that the algebra does not close for supergravity when we use the 1.5 order formalism. We need to start with a correct formulation of the first order formalism, and then get the second order formulation using the master equation for supergravity too.
\item
Of course this is a huge calculation in supergravity, though it is a simple one in the WZ model.
\een

\refstepcounter{orange}
{\bf \theorange}.\;
{\bf This is not at all surprising:} From the general form of the Master Equation in  (\ref{1piequation}) we can see that there is a $\d$ that is nilpotent of the form:

\be
\d=    \int d^4 x \; 
\sum_{i=1}^N
\lt \{
\fr{\d \cG }{\d {\rm \f}^i} \fr{\d  }{\d \widetilde{\f}_i} 
+
\fr{\d \cG}{\d \widetilde{\f}_i}  \fr{\d   }{\d {\rm \f}^i} 
\rt \}
\ee
and it follows from  (\ref{1piequation}) that 
\be
\d^2 =0
\ee
These are both identities, not merely equations that are true assuming some equations of motion. However if we remove a term, which is what happens in the 1.5 formalism, we get:
\be
\cM_{\rm Term\;Missing}=     \sum_i^{N-1}
\int d^4 x \; \fr{\d \cG }{\d {\rm \f}^i} \fr{\d \cG}{\d \widetilde{\f}_i}  \neq 0
\la{1piequation2}
;\;
\eb
\d_{\rm Term\;Missing}=    \int d^4 x \; 
\sum_{i=1}^{N-1} 
\lt \{
\fr{\d \cG }{\d {\rm \f}^i} \fr{\d  }{\d \widetilde{\f}_i} 
+
\fr{\d \cG}{\d \widetilde{\f}_i}  \fr{\d   }{\d {\rm \f}^i} 
\rt \}\;{\rm Not\; nilpotent}
\ee
In this case we can be sure that we do not get zero for the first and we do not get nilpotence for the second.
We saw explicitly above that the nilpotence does not arise even if we substitute the value of the field equation for the field.

\refstepcounter{orange}
{\bf \theorange}.\;
\la{whatnowpar}
{\bf BRS Cohomology and Supergravity:} This problem stands in the way of computing the BRS cohomology of supergravity  in a convincing way. 
For me, this problem arose from \ci{ramble}, when  I gradually realized that I could not find out how  to write down the nilpotent transformations. The methods in \ci{Dixon:1991wi}  require identities for the nilpotence. However as shown above in paragraph \ref{wrongway}, using the 1.5 formalism, one gets  BRS operators that are not nilpotent at all, not even when using the equations of motion.

\refstepcounter{orange}
{\bf \theorange}.\;
\la{spinconisYM}
{\bf  It seems rather easy to close the first order algebra here.}  As is well known, the Einstein curvature scalar $R$ looks like a YM curvature made from the spin connection, contracted with vierbeins.  So the way to make this supersymmetric is evident. We add a new auxiliary spinor field $\W^{ab\a},\ov\W^{ab\dot\a}$ 
and a new auxiliary scalar  $D^{ab}$, 
and with   $w_{\m}^{ab}$   and the ghost  $\r^{ab}$,  these make a full super--YM multiplet.   This multiplet is analogous to the usual super--YM multiplet, except that the gauge group is the non-compact Lorentz group. The details of the transformations need careful work.  This new multiplet raises many issues. One of the problems is that this makes it very tempting to add the usual quadratic super-Yang Mills type action for the new multiplet, but  that would mean that the new super--YM multiplet no longer consists of auxiliary fields.  They would become propagating and dynamical.

\begin{center}
 {\bf Acknowledgments}
\end{center}
\vspace{.1cm}

  I thank  Doug Baxter, Carlo Becchi,   Friedemann Brandt, James Dodd, Mike Duff,     Pierre Ramond,  Peter Scharbach,    Kelly Stelle,    J.C. Taylor, and Peter West for recent stimulating correspondence and conversations.

\vspace{1cm}\tiny\numberhere \hspace{.2cm}\articlenumber\\ \today \hourandminute
\end{document}